\documentclass[11pt]{article}

\usepackage{thmtools}
\usepackage{thm-restate}
\usepackage{cleveref}

\usepackage{mathrsfs}           
\usepackage{algorithm}
\usepackage{algorithmic}

\usepackage{enumerate}

\usepackage{amsmath,amssymb}    
\usepackage{verbatim}           
\usepackage{xspace}             
\usepackage{graphicx,float}     
\usepackage{ifthen,calc}        
\usepackage{textcomp}           
\usepackage{fancybox}           
\usepackage{hhline}             
\usepackage{float}              

\usepackage[rflt]{floatflt}     
\usepackage[small,compact]{titlesec}
\usepackage{subfigure}

\usepackage{color}
\definecolor{ForestGreen}{rgb}{0.1333,0.5451,0.1333}
\usepackage[letterpaper,
            colorlinks,linkcolor=ForestGreen,citecolor=ForestGreen,
            backref,
            bookmarks,bookmarksopen,bookmarksnumbered]
           {hyperref}
\usepackage{thumbpdf}

\usepackage[margin=1.01in]{geometry}

\newtheorem{theorem}{Theorem}[section]
\newtheorem{claim}[theorem]{Claim}

\newtheorem{corollary}[theorem]{Corollary}
\newtheorem{definition}[theorem]{Definition}
\newtheorem{conjecture}[theorem]{Conjecture}

\newtheorem{lemma}[theorem]{Lemma}

\newcommand{\Proof}[0]{\smallskip\noindent\textit{\textbf{Proof:}}\quad}
\newcommand{\Proofsketch}[0]{\smallskip\noindent\textit{\textbf{Sketch of proof:}}\quad}
\newcommand{\Proofof}[1]{\smallskip\noindent\textit{\textbf{Proof of #1:}}\quad}

\newcommand{\Oh}{\ensuremath{\mathcal{O}}}
\newcommand{\Ohstar}{\ensuremath{\Oh^\star}}

\newcommand{\QED}[0]{\hfill\ensuremath{\blacksquare}\medspace\\}

\floatstyle{ruled}
\newfloat{algo}{tbp}{lop}
\floatname{algo}{Algorithm}

\usepackage{float}
\floatstyle{plain}\newfloat{myfig}{t}{figs}[section]
\floatname{myfig}{\textsc{Figure}}
\floatstyle{plain}\newfloat{myalg}{H}{algs}[section]
\floatname{myalg}{}
\newcommand{\poly}[1]{\text{poly}(#1)}

\usepackage{standalone}
\usepackage{tikz}
\usetikzlibrary{arrows}
\usetikzlibrary{patterns}

\usepackage{url}

\newcommand{\CA}[0]{\textsc{Channel Assignment}\xspace}  
\newcommand{\SI}[0]{\textsc{Subgraph Isomorphism}\xspace}
\newcommand{\GH}[0]{\textsc{Graph Homomorphism}\xspace}
\newcommand{\SIp}[1]{#1-\textsc{Subgraph Isomorphism}\xspace}
\newcommand{\TSAT}[0]{$3$-\textsc{SAT}\xspace}
\newcommand{\eps}{\varepsilon}
\newcommand{\Var}{{\mathrm{Var}}}
\newcommand{\val}{{\mathrm{val}}}
\newcommand{\true}{{\bf{true}}}
\newcommand{\false}{{\bf{false}}}
\newcommand{\Clauses}{{\mathrm{Clauses}}}

\def\cqedsymbol{\ifmmode$\lrcorner$\else{\unskip\nobreak\hfil
\penalty50\hskip1em\null\nobreak\hfil$\lrcorner$
\parfillskip=0pt\finalhyphendemerits=0\endgraf}\fi}

\newcommand{\defproblemu}[3]{
  \vspace{2mm}
  \vspace{1mm}
\noindent\fbox{
  \begin{minipage}{0.95\textwidth}
  #1 \\
  {\bf{Input:}} #2  \\
  {\bf{Question:}} #3
  \end{minipage}
  }
  \vspace{2mm}
}

\begin{document}

\begin{titlepage}
\def\thepage{}
\thispagestyle{empty}

\title{The Hardness of Subgraph Isomorphism}

\author{Marek Cygan\footnote{Institute of Informatics, University of Warsaw, \texttt{cygan@mimuw.edu.pl}}
\and Jakub Pachocki\footnote{Carnegie Mellon University, \texttt{pachocki@cs.cmu.edu}}
\and Arkadiusz Soca\l{}a\footnote{Institute of Informatics, University of Warsaw, \texttt{a.socala@mimuw.edu.pl}}}

\maketitle

\abstract{
Subgraph Isomorphism is a very basic graph problem, where given two graphs $G$ and $H$ 
one is to check whether $G$ is a subgraph of $H$.
Despite its simple definition, the Subgraph Isomorphism problem turns
out to be very broad, as it generalizes problems such as Clique, $r$-Coloring, Hamiltonicity, Set Packing and Bandwidth.
However, for all of the mentioned problems $2^{\Oh(n)}$ time algorithms exist, so a natural and frequently
asked question in the past was whether there exists a $2^{\Oh(n)}$ time algorithm for Subgraph Isomorphism.
In the monograph of Fomin and Kratsch [Springer'10] this question is highlighted as an open problem, among few others.

Our main result is a reduction from \TSAT, producing a subexponential number of sublinear instances of the Subgraph Isomorphism
problem. In particular, our reduction implies a $2^{\Omega(n \sqrt{\log n})}$ lower bound for Subgraph Isomorphism
under the Exponential Time Hypothesis.
This shows that there exist classes of graphs that are strictly harder to embed than cliques or Hamiltonian cycles.

The core of our reduction consists of two steps. First, we preprocess and pack variables and clauses of a \TSAT formula
into groups of logarithmic size. However, the grouping is not arbitrary, since as a result we obtain only a limited interaction
between the groups.
In the second step, we overcome the technical hardness of encoding evaluations as permutations by
a simple, yet fruitful scheme of guessing the sizes of preimages of an arbitrary mapping,
reducing the case of arbitrary mapping to bijections. In fact, when applying this step to a recent independent
  result of Fomin et al.[CoRR abs/1502.05447 (2015)], who showed hardness of Graph Homomorphism,
we can transfer their hardness result to Subgraph Isomorphism, implying a nearly tight lower bound of $2^{\Omega(n \log n / \log \log n)}$.
}
\end{titlepage}

\section{Introduction}

Perhaps the most basic relation between graphs is that of being a subgraph.
We say that $G$ is a subgraph of $H$ if one can remove some edges and vertices
of $H$, so that what remains is isomorphic to $G$.
Formally, the question of one graph being a subgraph of another
is the base of the \SI problem.

\defproblemu{\SI}{ undirected graphs $G$, $H$. } {is $G$ a subgraph of $H$, i.e., does there exist an injective function $g : V(G) \to V(H)$,
such that for each edge $uv \in E(G)$ we have $g(u)g(v) \in E(H)$.}

\SI is an important and very general question, having the form of a pattern matching -- 
we will call $G$ the \emph{pattern graph} and $H$ the \emph{host graph}.
Observe that several flagship graph problems can be viewed as instances of \SI:
\begin{itemize}
  \item {\sc Hamiltonicity}(G): is $C_n$ (a cycle with $n$ vertices) a subgraph of $G$?
  \item {\sc Clique}(G,k): is $K_k$ a subgraph of $G$?
  \item {\sc 3-Coloring}(G) : is $G$ a subgraph of $K_{n,n,n}$, a tripartite graph with $n$ vertices in each of its three independent sets?
  \item {\sc VertexCover}(G,k) : is $G$ a subgraph of $H$, $H$ being a full join between a clique of size $k$ and an independent set of size $n-k$?
\end{itemize}
One can continue showing the richness of \SI by simple linear reductions from {\sc Bandwidth}, {\sc Set Packing} and several other problems.

All of the mentioned problems are NP-complete, and the best known algorithms
for all the listed special cases work in exponential time.
In fact, all those problems are well-studied from the exact exponential algorithms perspective~\cite{3-col, hamiltonicity, col, mis, bw},
where the goal is to obtain an algorithm of running time $\Oh(c^n)$ for smallest possible value of $c$.
Furthermore, the \SI problem was very extensively studied from the viewpoint of fixed parameter tractability, see~\cite{michal} for a discussion of 19 different possible parametrizations.
All the mentioned special cases of \SI admit $\Oh(c^n)$ time algorithms, by using either branching,
inclusion-exclusion principle or dynamic programming.
On the other hand, a simple exhaustive search for the \SI problem
-- numerating all possible mappings from the pattern graph to the host graph --
runs in $2^{\Oh(n \log n)}$ time, 
where $n$ is the total number of vertices of the host graph and pattern graph.

Therefore, a natural question is whether \SI admits an $\Oh(c^n)$ time algorithm.
This was repeatedly posed as an open problem~\cite{bedlewo, counting-homo, dagstuhl-1, fomin-homo, dagstuhl-2}. 
In particular, Fomin and Kratsch in their monograph~\cite{fomin-kratsch} put the existence of $\Oh(c^n)$ time algorithm
for \SI among the few questions in the open problems section.

\paragraph{Our results and techniques}

Our main result is a reduction
which transforms a \TSAT formula into a subexponential number
of sublinear instances of the \SI problem.
This implies that a $\Oh(c^n)$ time algorithm for \SI
would imply a subexponential algorithm for \TSAT,
thus refuting the Exponential Time Hypothesis of Impagliazzo, Paturi and Zane~\cite{eth1, eth2}.
The Exponential Time Hypothesis 
is an established assumption; several interesting lower bounds have been found under this conjecture (see~\cite{eth-survey} for a survey).

\begin{theorem}
\label{thm:main}
There is no algorithm which solves \SI in $2^{o\left(n \sqrt{\log n}\right)}$ time,
unless the Exponential Time Hypothesis fails.
\end{theorem}

Our reduction can be broken into three steps:
\begin{itemize}
\item First, in Section~\ref{sec:grouping}, we preprocess the given \TSAT formula
and pack its variables and clauses into groups of logarithmic size.
Importantly, we ensure that there is only a limited 
interaction between the groups by marking variables with colors -- applying further
steps of the reduction for an arbitrary grouping would not yield a superexponential
lower bound for \SI.
\item Next, in Section~\ref{sec:reduction}, we use the packing to create 
$2^{\Oh(n / \log n)}$ smaller instances of a variant of the \SI problem, 
where additionally vertices and edges have colors which have to 
be preserved by the mapping.
This proves that the color variant of \SI admits
a tight lower bound of $2^{\Omega(n \log n)}$ under the Exponential Time Hypothesis.
In this step, we use a simple technique of guessing preimage sizes,
which allows us to circumvent the usual technical difficulties of 
encoding valuations by permutations.
\item Finally, in Section~\ref{sec:slog} we reduce the color version of \SI
to the original variant, incurring an $\Oh(\sqrt{\log n})$ increase
in the instance size.
\end{itemize}

We would like to note that very recently and independently,
Fomin et al.~\cite{fomin-homo}, in an unpublished work,
proved that under the Exponential Time Hypothesis
there is no $2^{o(n \log h / \log \log h)}$
time algorithm for a related problem called \GH.
\GH has a similar definition to \SI,
except that the mapping is not constrained
to be injective (i.e., in a homomorphism
many vertices of the pattern graph may be mapped to the same vertex of the host graph).
One could think that \GH is a harder problem than \SI,
as for example in~\cite{counting-homo} Amini et al. have shown
that counting subgraphs can be reduced to counting homomorphisms.
In fact, Fomin et al.~\cite{fomin-homo} in their work about \GH
mention the question about \SI as an open problem.

\begin{theorem}{\cite{fomin-homo}}
\label{thm:fomin-homo}
    There is no algorithm which solves \GH in $2^{o(n \log h / \log \log h)}$ time, where
    $h = \Oh(\poly{n})$ is the size of the host graph and $n$ is the size of the pattern graph, unless the Exponential Time Hypothesis fails.
\end{theorem}

In Section~\ref{sec:homo} we prove that
by applying our simple scheme of guessing preimage sizes,
one can transform an instance of \GH into an exponential number of
instances of \SI.

\begin{theorem}
\label{thm:red-homo}
Given an instance $(G,H)$ of \GH one can in $\Oh(2^{n} \poly{n})$
time create $2^n$ instances of \SI with $n$ vertices, where $n = |V(G)| + |V(H)|$,
such that $(G,H)$ is a yes-instance iff at least one of the created
instances of \SI is yes-instance. 
\end{theorem}

Note that Theorem~\ref{thm:red-homo}, when combined with the lower
bound of Fomin et al. quoted in Theorem~\ref{thm:fomin-homo}, implies a stronger lower bound for \SI.

\begin{corollary}
There is no algorithm which solves \SI in $2^{o\left(n \log n / \log \log n\right)}$ time,
unless the Exponential Time Hypothesis fails.
\end{corollary}

\section{Preliminaries}

\paragraph{Notation}
We use the convention $[k] = \{0,\ldots,k-1\}$.
All the graphs used in this article are undirected, 
however in edge colored graphs there might be several
parallel edges between the same pair of vertices.
We use standard graph notation -- for an undirected graph $G$, by $V(G)$ we denote the set of
vertices of $G$, whereas by $E(G)$ we denote the set of edges of $G$.

For a CNF-SAT formula $\varphi$ let $\Var(\varphi)$ be
the set of variables of $\varphi$, whereas $\Clauses(\varphi)$
is the set of clauses of $\varphi$.

By saying that two instances $I$, $I'$ of some decision problems $P$ and $Q$, respectively,
are equivalent, we mean that $I$ is a yes-instance of the problem $P$ iff $I'$ is a yes instance
of the problem $Q$. In particular two formulas are equivalent iff they are either 
none of both of them are satisfiable.

To simplify the reduction we use the standard method
of transforming a \TSAT formula into an equivalent
formula with exactly three different variables in each clause
and each variable occurring in at most $4$ clauses.

\begin{lemma}{\cite{sat-transformation}}
\label{lem_transformation}
Given a \TSAT formula $\varphi$ with $m$ clauses one can transform
it in polynomial time into a formula $\varphi'$ with $\Oh(m)$ 
variables and $\Oh(m)$ clauses, such that $\varphi'$ is
satisfiable iff $\varphi'$ is satisfiable, and
moreover each clause of $\varphi'$ contains exactly three
variables and each variable occurs in at most $4$ clauses
of $\varphi'$.
\end{lemma}

\paragraph{Exponential Time Hypothesis}

The Exponential Time Hypothesis, introduced by Impagliazzo, Paturi and Zane~\cite{eth1, eth2},
states that it is impossible to solve \TSAT in time subexponential in the number of variables.
Note that the $\Ohstar()$ notation suppresses polynomial factors.

\begin{conjecture}[Exponential Time Hypothesis~\cite{eth2}]
There exists a constant $c > 0$, such that 
there is no algorithm solving \TSAT in time $\Ohstar(2^{cn})$.
\end{conjecture}

One of the reasons why the Exponential Time Hypothesis became a robust tool
for proving lower bounds is the Sparsification Lemma, which allows 
to reduce the number of clauses in a formula to be linear in the number of variables.

\begin{lemma}[Sparsification Lemma~\cite{eth1}]
\label{lem_sparsification}
For each $\eps > 0$ there exist a constants $c_{\eps}$,
such that any \TSAT formula $\varphi$ with $n$ variables
can be expressed as $\varphi = \vee_{i=1}^t \psi_i$, where $t \le 2^{\eps n}$ 
and each $\psi_i$ is a \TSAT formula with the same variable set
as $\varphi$, but contains at most $c_{\eps}n$ clauses.
Moreover, this disjunction can be computed in time $\Ohstar(2^{\eps n})$.
\end{lemma}

\section{Overview}

We define the \emph{size} of a \SI instance to be the total number of vertices in the pattern and host graphs.

\begin{definition}
    We define the \SIp{$(c, t)$} problem as a generalization of \SI where every vertex of the pattern and host graphs is colored in one of $c$ colors, and every edge is colored in one of $t$ colors, and the mapping is restricted to preserving vertex and edge colors.
\end{definition}

In particular, \SI is the same as \SIp{$(1, 1)$}.
The pipeline of our lower bound consists of two steps.
First, in Lemma~\ref{lem_color_red}, 
given a \TSAT formula with $n$ variables
we construct a set of $2^{\Oh(n / \log n)}$ instances of \SIp{$(\Oh(1), \Oh(\log n))$}
of $\Oh(n / \log n)$ size each. 
Note that the number of vertex colors is constant, whereas the number of edge colors is logarithmic.
In the second step (Lemma~\ref{lem_slog}) we reduce to the original variant of \SI,
with an additional increase in the instance size by a factor of $\Oh(\sqrt{\log n})$,
leading to a final size of $\Oh(n / \sqrt{\log n})$, which is sublinear.

\begin{lemma}
\label{lem_color_red}
Given a \TSAT formula $\varphi$ with $n$ variables, 
where each variable occurs in at most $4$ clauses
and each clause involves exactly three variables,
one can in $2^{\Oh(n / \log n)}$ time create a set $\mathcal{S}$ of $2^{\Oh(n / \log n)}$ instances
of \SIp{$(\Oh(1), \Oh(\log n))$} of size $\Oh(n / \log n)$,
such that $\varphi$ is satisfiable iff any instance in $\mathcal{S}$ is satisfiable,
and the host graph and the pattern graph have the same number of vertices for every instance in $\mathcal{S}$.
\end{lemma}

\begin{lemma}
\label{lem_slog}
An instance of \SIp{$(c, t)$}, where the host graph and the pattern graph
have the same number of vertices, can be reduced to an
equivalent instance of \SI with $\Oh(c\sqrt{t})$ times more vertices.
\end{lemma}

Having the two lemmas above, which we prove in the remainder of this paper,
we can prove Theorem~\ref{thm:main}.

\Proofof{Theorem~\ref{thm:main}}
Assume that a $2^{o(n \sqrt{\log n})}$ time algorithm
exists for the \SI problem, where $n = |V(G)| + |V(H)|$.
For a given $\eps > 0$, we show an algorithm solving
a given \TSAT formula $\varphi$ with $n$ variables in time $\Ohstar(2^{3\eps n})$,
leading to a contradiction with the Exponential Time Hypothesis.

First, we sparsify the formula using Lemma~\ref{lem_sparsification}
to obtain $\Ohstar(2^{\eps n})$ formulas $\psi_i$, each with
$n$ variables and $\Oh(n)$ clauses (where the hidden constant depends on $\eps$).
Consider each $\psi_i$ independently.
For a fixed $\psi_i$, we use Lemma~\ref{lem_transformation} to
obtain an equivalent formula $\psi_i'$ with $\Oh(n)$ variables and clauses,
with the additional property that each clause involves exactly three variables
and each variable appears in at most $4$ clauses.
Consequently, the prerequisites of Lemma~\ref{lem_color_red} are satisfied,
and in $2^{\Oh(n / \log n)}$ time we can obtain
a corresponding set $\mathcal{S}$ of $2^{\Oh(n / \log n)}$ instances of \SIp{$(\Oh(1), \Oh(\log n))$} of size $\Oh(n / \log n)$ each.
Next, we apply Lemma~\ref{lem_slog} to transform each instance in $\mathcal{S}$ into an instance
of \SI of size $\Oh(n / \sqrt{\log n})$, obtaining the set $\mathcal{S}'$.
Finally, we apply the hypothetical $2^{o(n \sqrt{\log n})}$-time 
algorithm to the instances in $\mathcal{S}'$, leading to $2^{\Oh(n / \log n)}2^{o(n)} = 2^{o(n)}$ running time.
Note that the total running time is $\Ohstar(2^{\eps n}) \cdot \Ohstar(2^{\eps n}) \cdot 2^{o(n)}$,
which is not more than $\Ohstar(2^{3\eps n})$, as promised, hence the theorem follows.
\QED

We prove Lemma~\ref{lem_color_red} in Section~\ref{sec:reduction}
and Lemma~\ref{lem_slog} in Section~\ref{sec:slog}.
However, before we describe the reduction,
in Section~\ref{sec:grouping} we present 
how to group clauses of a given \TSAT formula
in a way that allows a sublinear reduction to \SI.

\section{Grouping clauses}
\label{sec:grouping}

As we already mentioned, when proving superexponential lower bounds based on
the Exponential Time Hypothesis, we need to come up with a reduction
producing an instance of \SI of sublinear size. 
In this section we show how to preprocess a given \TSAT formula
and partition its clauses into groups of logarithmic size.
Our grouping is far from arbitrary, as we need to precisely
control the interactions between clauses sharing the same variables.

Before we arrive at our main structural lemma, we need a simple step
in which we assign colors to variables so that no clause contains
two variables of the same color and moreover the counts of variables in each color are balanced.
The proof of the following Lemma is contained in Appendix~\ref{app:grouping}.

\begin{lemma}
\label{lem_balanced_coloring} $(\spadesuit)$
Given an integer $k > 9$ and a $\TSAT$ formula $\varphi$ with $n$ variables, where each variable occurs in at most $4$ clauses,
we can color the variables of $\varphi$ in polynomial time using at most $k$ colors,
so that no more than $\lceil n / (k - 9) \rceil$ variables share the same color 
and no clause contains two variables of the same color.
\end{lemma}

Having Lemma~\ref{lem_balanced_coloring} we are ready to pack the clauses of a given \TSAT
formula into $2^k$ groups, which is the main structural insight in our reduction.
It is important that no two clauses from the same group contain variables of the same color.

\begin{lemma}
\label{lem_packing}
Given a $\TSAT$ formula $\varphi$ with $n \ge 16$ variables, such that 
each clause involves exactly three variables and each variable occurs in at most $4$ clauses,
one can in polynomial time construct: 
\begin{itemize}
    \item a coloring $l : \Var(\varphi) \to [k]$ of the variables in $\varphi$ into $k$ colors, 
    such that no two variables contained in a clause of $\varphi$ share the same color, and
    \item a packing $h : \Clauses(\varphi) \to [2^k]$ of the clauses into $2^k$
        groups indexed by $\{0, \ldots, 2^k - 1\}$, such that for any $i \in [2^k]$
    no two clauses that are mapped to $i$ contain variables of the same color,
\end{itemize}
where $k := \lceil \log n - \log \log n \rceil + 9$.
\end{lemma}

\Proof
    Let $l$ be the coloring guaranteed by Lemma \ref{lem_balanced_coloring}.
    We slightly overload the notation and by $l(C)$ denote the set
    of colors of variables in $C \in \Clauses(\varphi)$.

    We construct the packing $h$ in a greedy manner.
    Consider all the clauses of $\Clauses(\varphi)$ one by one in an arbitrary order.
    When a clause $C \in \Clauses(\varphi)$ is processed, we find
    any group $i \in [2^k]$, such that the set of colors of variables appearing in clauses already
    assigned to $i$ is disjoint from $l(C)$.
    If several such sets $i$ exist, we pick an arbitrary one and assign $h(C) := i$.

    It remains to prove that such an $i$ always exists for the value of $k$ as stated in the lemma.
    We prove this by contradiction: suppose that at some point, for some clause $C$,
    for every $i$ one of the colors in $l(C)$ is already present 
    in a clause already assigned to $i$.
    Let $m_{l(C)}$ be the number of clauses of $\varphi$ 
    containing at least one color from $l(C)$.
    As there are exactly $2^{k}$ groups,
    and we cannot assign $C$ to any of them, it means that 
    \begin{align}
    \label{ineq1}
    m_{l(C)} \ge 2^{k} \ge 512 n / \log n\,,
    \end{align}
    since each of the $2^{k}$ groups is blocked by a different clause
    containing at least one color from $l(C)$.

    On the other hand we have only $3$ colors
    in $l(C)$ and we know by
    Lemma \ref{lem_balanced_coloring},
    that no more than $\lceil n / (k - 9) \rceil$ variables are assigned to any color, 
    and by the upper bound on the frequency of each variable of $\varphi$
    we know that no variable occurs in more than $4$ clauses. 
    Consequently, the number of clauses having at least one common color with $C$
    is upper bounded by 
    \begin{align}
    \nonumber
    m_{l(C)} &
      \le 3 \cdot \lceil n / (k-9) \rceil \cdot 4
      \le 12 \cdot (n / (k-9) + 1) \\
    \nonumber &
      \le 12 \cdot (n / (\log n - \log \log n) + 1)
      \le 12 \cdot (2n / \log n + 1) \\
    \label{ineq2} &
      \le 12 \cdot (2n / \log n + 0.5 n / \log n)
      \le 30n / \log n\,,
    \end{align}
    where in the last two inequalities we have used that $\log n - \log \log n \ge 0.5 \log n$ and $n / \log n \ge 2$ for $n \ge 16$.
    Note that~(\ref{ineq2}) yields a contradiction with~(\ref{ineq1}), and the lemma follows.
\QED

\section{From \TSAT to \SI with colors}
\label{sec:reduction}

The technical crux of our result is a method of encoding information in permutations -- mappings
from the pattern graph to the host graph.
The intuition behind this technique is that the number
of permutations of an $n$ element set is
$n! = 2^{\Theta(n\log n)}$ and therefore a single permutation
carries $\Theta(n\log n)$ bits of information.
This means that from the information-theoretic perspective
if should be possible to encode an assignment of Boolean values
to $n$ variables using a permutation of $\Oh(n /\log n)$ elements.

Every element in a permutation is
responsible for encoding some number of bits,
forming what we call a \emph{pack} of bits.
We do not restrict ourselves to packs of constant size,
but each pack we create is of size no greater than logarithmic.
The position of an element in a permutation should
uniquely determine the values of all the bits from its pack.
The problem is, however, that it in a permutation
no two elements can be mapped to the same position,
which potentially might make it impossible to assign
the same valuation to two different packs of bits.

Here, we present a new and simple way of circumventing
this obstacle by guessing the sizes of preimages
in a mapping corresponding to a satisfying assignment.
Less formally, what we do is replicate some positions 
and remove other ones, so that in some branch 
our guess will transform a mapping we had in mind
into a permutation.

We would like to note that encoding groups of bits
by a position in a permutation was already used
by Marx, Lokshtanov and Saurabh~\cite{slightly-superexponential} in the $k \times k$-{\sc Permutation Clique}
problem, as well as by Soca\l{}a~\cite{Socala15} in the lower bound for the \CA problem.
Both of these two reductions (especially Lemma~2.3 from~\cite{Socala15}) could be simplified 
when using our guessing preimage sizes approach, instead of a technical one-to-one reduction.

In the remainder of this section we prove Lemma~\ref{lem_color_red}, that is show
how to transform a \TSAT formula $\varphi$ into $2^{\Oh(n /\log n)}$
instances of \SIp{$(\Oh(1), \Oh(\log n))$} with $\Oh(n / \log n)$ vertices.
In order to do this we need to introduce notation for binary strings.
Assume for a moment, that $n$ is a power of two, i.e., $n = 2^k$ for $k \in \mathbb{N}$.
One can view elements in a permutation as integers between $0$ and $n-1$,
denoted as $[n]$, but also as a set of binary strings of length $k$
-- being the binary representations of numbers from $[n]$, denoted as $2^{[k]}$.
We will use the two conventions interchangeably and for this reason
we need the following notation regarding binary strings. 
Let $\mathcal{B} := \{0, 1\}^*$ be the set of all binary strings.
and $\mathcal{B}_k := \{0, 1\}^k$ be the set of binary strings of size exactly $k$.
For a binary string $s$, let $|s|$ be its length.
We denote the $i$-th digit (starting from $0$) of a binary string $s$ as $s_i$.

\Proofof{Lemma \ref{lem_color_red}}
Assume we are given a formula $\varphi$ with $n$ variables, such that each
clause involves exactly three variables and each variable appears in at most $4$
clauses.
Define $k := \lceil \log n - \log \log n \rceil + 9$.
We prove that solving $\varphi$ can be reduced to solving less than $2^{2^{k+1}} = 2^{\Oh(n / \log n)}$ instances of \SIp{$(3, k)$}, with vertex colored denoted as red, green and blue, and
edge colors denoted by $[k]$, where the number of vertices of both the pattern and host graph equals
\begin{align*}
    2^k + 8\cdot{k \choose 3} + 1 = \Oh(n / \log n)\,.
\end{align*}

\noindent\textbf{Satisfying assignment gadget.}\\
The assignment gadget $G$ consists of a path on $8\cdot{k \choose 3}$ red vertices, with a single green vertex appended at one end.
The red vertices will be uniquely identifiable based on the distance from the green vertex.
Each red vertex will correspond to a choice of $3$ distinct indices from $[k]$ and an assignment of binary values to each of them:
\begin{align*}
    (i_1, i_2, i_3, b_1, b_2, b_3) &\in [k]^3 \times \mathcal{B}_3\,,\\
                             i_1 < i_2 < i_3\,.
\end{align*}
Intuitively, an edge between one of the clause vertices and a red vertex will indicate that `in this pack of clause valuations, the variables at positions $i_1, i_2, i_3$ are \emph{not} assigned values $b_1, b_2, b_3$ at the same time'.
All edges in $G$ are of color $0$.
\\\\\noindent\textbf{Pattern graph construction.}\\
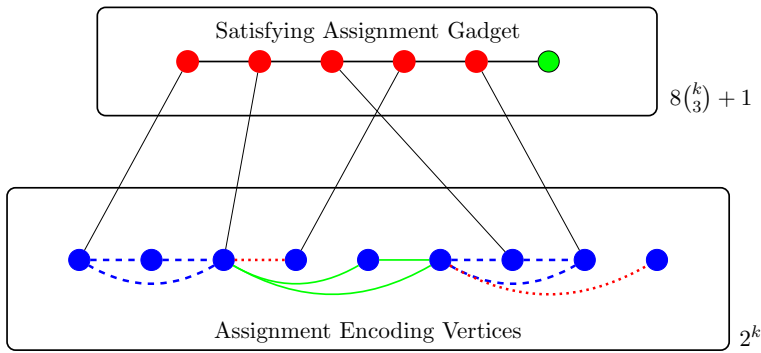
\begin{figure}
	\vspace{-2.3cm}
	\begin{center}
		\scalebox{0.8}{\begin{tikzpicture}[scale=0.6]
\tikzstyle{every node}=[circle, draw, color=black, fill=black, inner sep=0pt, minimum width=10pt]

\draw [thick](-2.5,0.5) node [color=red] (v14) {} -- (-0.5,0.5) node [color=red] (v12) {} -- (1.5,0.5) node [color=red] (v15) {} -- (3.5,0.5) node [color=red] (v13) {} -- (5.5,0.5) node [color=red] (v1) {};
\node [fill=green] (v2) at (7.5,0.5) {};
\draw [thick] (v1) edge (v2);

\node [color=blue] (v9) at (-5.5,-5) {};
\node [color=blue] (v10) at (-3.5,-5) {};
\node [color=blue] (v3) at (-1.5,-5) {};
\node [color=blue] (v4) at (0.5,-5) {};
\node [color=blue] (v8) at (2.5,-5) {};
\node [color=blue] (v5) at (4.5,-5) {};
\node [color=blue] (v6) at (6.5,-5) {};
\node [color=blue] (v7) at (8.5,-5) {};
\node [color=blue] (v11) at (10.5,-5) {}; 
\draw [very thick, color=blue, dashed] (v5) edge (v6);
\draw [very thick, color=blue, dashed] (v6) edge (v7);
\draw [very thick, color=blue, dashed, bend right] (v5) edge (v7);
\draw [thick, color=green, bend right] (v3) edge (v8);
\draw [thick, color=green, bend right] (v3) edge (v5);
\draw [thick, color=green] (v8) edge (v5);
\draw [very thick, color=blue, dashed] (v9) edge (v10);
\draw [very thick, color=red, bend right, dotted] (v5) edge (v11);
\draw [very thick, color=red, dotted] (v3) edge (v4);
\draw [very thick, color=blue, dashed] (v10) edge (v3);
\draw [very thick, color=blue, dashed, bend left] (v3) edge (v9);

\draw [rounded corners, thick] (-5,2) rectangle (10.5,-1);
\draw [rounded corners, thick] (-7.5,-3) rectangle (12.5,-7.5);

\node [draw=none, fill=none] at (2.5,1.30) {
  Satisfying Assignment Gadget};
\node [draw=none, fill=none] at (2.5,-7) {
  Assignment Encoding Vertices};
\node [draw=none, fill=none] at (12,-0.5) {
  $8 {k \choose 3} + 1$};
\node [draw=none, fill=none] at (13.1,-7.1) {
  $2^k$};

\draw (v4) edge (v13);
\draw (v12) edge (v3);
\draw (v15) edge (v6);
\draw (v7) edge (v1);
\draw (v14) edge (v9);
\end{tikzpicture}}
	\end{center}
	\vspace{-2.5cm}
	\caption{A simplified view of the pattern graph.}
	\label{fig:pattern}
\end{figure}
The pattern graph will be constant across all the created instances.
The pattern graph $P$ consists of $2^k$ blue vertices corresponding to packs of clauses and a copy of the satisfying assignment gadget $G$.
First, find the coloring $l$ and packing $h$ guaranteed by Lemma $\ref{lem_packing}$.
We associate each blue vertex of $H$ with a different group in the image of $h$.

For every variable $x$ in $S$ and every two distinct clauses $C_1, C_2$ containing $x$, we add an edge of color $l(x)$ between the blue vertices corresponding to $h(C_1)$ and $h(C_2)$.
Intuitively, these edges signify that $x$ has to have a consistent valuation when choosing valuations of variables in packs containing $C_1$ and $C_2$.

Additionally, for every clause $C$ in $\varphi$ we add an edge of color $0$ between $h(C)$ (i.e., the pack containing $C$) and the red vertex $(i_1, i_2, i_3, b_1, b_2, b_3)$, where $i_1 < i_2 < i_3$ are the colors of variables in $C$ and $(b_1, b_2, b_3)$ is their only valuation that does not satisfy $C$.
\\\\\noindent\textbf{Host graph construction.}\\
We will generate a different host graph for every sequence of preimage sizes of the valuations of the groups.
Fix a sequence $s_0, s_1, \ldots, s_{2^k - 1}$, such that $s_i \geq 0$ for all $i$ and $\sum s_i = 2^k$.

The number of possible such sequences $s$ is
\begin{align*}
    {2^{k + 1} - 1} \choose {2^k - 1} &\leq 2^{2^{k+1}}.\\
\end{align*}

The host graph $H_s$ consists of $2^k$ blue vertices corresponding to valuations of the groups of clauses and a copy of the satisfying assignment gadget $G$.
For the binary string of length $k$ corresponding to $i \in [2^k]$, we generate $s_i$ vertices corresponding to it.

For $j \in \mathbb{Z}_k$, we join two blue vertices $u, v$ in $H$ with an edge of color $j$ iff $u_j = v_j$, that is
iff the $j$-th bit in both strings is the same.
Intuitively, lack of an edge of color $j$ between two blue vertices $u, v$ in $H$
disallows assigning two packs of clauses to vertices $u$ and $v$ when the variable of color $j$ in
both packs is the same, as it would lead to inconsistent valuation.

For every blue vertex $u$ in $H$ and red vertex $v = (i_1, i_2, i_3, b_1, b_2, b_3)$ in $G$, we connect $u$ and $v$ with an edge of color $0$ iff $u_{i_j} \neq b_j$ for some $j \in \mathbb{Z}_3$.
Less formally, lack of an edge between a blue vertex $u$ and a red vertex $v = (i_1, i_2, i_3, b_1, b_2, b_3)$ means that 
a pack of clauses can be assigned to $u$, only if the valuation corresponding to the bit string associated with $u$ only if there is no clause
such that assigning values $b_1,b_2,b_3$ to variables of colors $i_1,i_2,i_3$, respectively, would cause come clause from the pack to be unsatisfied.
\\\\\noindent\textbf{Proof of correctness \TSAT.}\\
As the construction can be carried out in polynomial time per instance and both the host and pattern graphs have
$\Oh(n / \log n)$ vertices as promised, it remains to prove that $\varphi$ is satisfiable iff for some instance the pattern graph $P$ is a subgraph of the host graph $H_s$.

\begin{claim}
\label{claim:1}
If $\varphi$ is satisfiable, then for some sequence of preimage sizes $s$, $P$ is a subgraph of $H_s$.
\end{claim}

\Proof
First, assume that $\varphi$ is satisfiable and let $\val : \Var(\varphi) \to \{\true, \false\}$
be a satisfying assignment.
We construct a mapping $g : V(P) \to V(H)$ as follows. 
For a group $i \in [2^k]$ let $\Var_i$ be the set of variables occurring
in all the clauses assigned to $i$ by the packing $h$.
If any colors do not occur in $l(\Var_i)$, add arbitrary variables to $\Var_i$
so that $l(\Var_i) = [k]$.
Define $f(i) = \val|_{\Var_i}$, i.e.,
the bit string representing valuation of variables from $\Var_i$ by $\val$.
Let $s$ be the sequence of preimage sizes of $f$.
A bijection $g$ corresponding to $f$ exists between the blue vertices of $P$ and $H_s$.
We extend $g$ to all the vertices of $P$ by mapping
each vertex of the satisfying assignment gadget $G$ in $P$ 
to its corresponding copy in $H_s$, obtaining a bijection $g' : V(P) \to V(H)$.

It remains to check that $b'$ preserves all the edges.
Clearly, the edges within the satisfying assignment gadget $G$ are preserved.
Consider any edge of color $c \in [k]$ in the pattern graph between two blue vertices
$u,v$, corresponding to groups $i$ and $j$.
By construction, this means that the packs $h^{-1}(i)$ and $h^{-1}(j)$
share a variable of color $c$, which means 
that by the definition of $f$ the bit strings $f(i)$ and $f(j)$ assign the same
value to the index corresponding to this variable.
As $g$ extends $f$,
we have $g(i) = f(i)$ and $g(j) = f(j)$, 
hence the bit strings corresponding to $g'(u)$ and $g'(v)$
have the same value on the $c$-th position, hence by construction
of the host graph $g'(u)$ and $g'(v)$ are connected by an edge of color $c$.
Finally, we inspect the edges between blue vertices and red vertices.
Consider a blue vertex $u$ associated with a set $A \subseteq [k]$,
which is connected to some red vertex $v=(i_1,i_2,i_3,b_1,b_2,b_3)$,
because of a clause $C \in h^{-1}(A)$.
As $\val$ is a satisfying assignment and $g$ extends 
bit strings assigned by $f$, we infer that the vertex $g(u)$
is connected to the red vertex $v$.
Consequently, $P$ is a subgraph of $H_s$ witnessed by the mapping $g'$.
\cqedsymbol

In Appendix~\ref{app:reduction} we prove the following claim.

\begin{claim}
\label{claim:2} $(\spadesuit)$
If for any $s$ it holds that $P$ is a subgraph of $H_s$, then $\varphi$ is satisfiable.
\end{claim}

Claims~\ref{claim:1} and~\ref{claim:2} prove equivalence of the formula $\varphi$
and created instances of \SIp{$(3,k)$}, hence the proof of Lemma~\ref{lem_color_red} follows.
\QED

We would like to note that Lemma~\ref{lem_color_red} implies
a tight bound for the auxiliary version of \SI with colors,
even in the case when the number of vertex colors is constant
and the number of edge colors is logarithmic.

\begin{corollary}
There is no $2^{o(n \log n)}$ time algorithm for the \SIp{$(\Oh(1),\Oh(\log n))$}
problem, unless the Exponential Time Hypothesis fails.
\end{corollary}

\section{Removing the colors}
\label{sec:slog}

In this section we prove Lemma~\ref{lem_slog},
first by showing how to remove colors from edges,
and next by removing colors from vertices.
Due to space constraints, we only sketch the
constructions, and the formal proof of the equivalence
of created instances is deferred to Appendix~\ref{app:removing_colors}.

\begin{lemma}
    \label{lem_edge_red} $(\spadesuit)$
    An instance $(G, H)$ of \SIp{$(c,t)$}
    such that $|V(G)| = |V(H)|$
    can be reduced to an instance $(G', H')$
    of \SIp{$(c + 1, 1)$} with $\Oh(\sqrt{t})$ times
    more vertices
    such that $|V(G')| = |V(H')|$.
\end{lemma}
\newcommand{\tprimhalf}{
	\ensuremath{\left\lceil \sqrt{t} \right\rceil}
}
\newcommand{\tprim}{
	\ensuremath{2\tprimhalf}
}
\begin{figure}
	\begin{center}
		\scalebox{0.5}{\begin{tikzpicture}[scale=0.4]
\tikzstyle{every node}=[circle, draw, color=yellow, fill=yellow, inner sep=0pt, minimum width=15pt]
\node[color=blue, fill=blue] (z1) at (-4.5,7.5) {};
\node[color=blue, fill=blue] (z2) at (-12.5,-4.5) {};
\node[color=black, fill=green] (z3) at (2.5,-4.5) {};
\draw[color=red, very thick, dotted] (z1) edge (z2);
\draw[color=green, very thick]  (z1) edge (z3);
\draw[color=blue, very thick, bend left, dashed]  (z2) edge (z3);
\draw[color=green, very thick, bend right]  (z2) edge (z3);

\node [color=black, preaction={fill, yellow}, pattern=grid] (u1) at (7.5,-1) {};
\node [color=black, preaction={fill, yellow}, pattern=grid] (u2) at (9,-2) {};
\node [color=black, preaction={fill, yellow}, pattern=grid] (u3) at (10.5,-3) {};
\node [color=black, preaction={fill, yellow}, pattern=grid] (u4) at (12,-4) {};
\node [color=black, preaction={fill, yellow}, pattern=grid] (u5) at (13.5,-5) {};
\node [fill=blue, color=blue] (u0) at (9,-4.5) {};
\node [color=black, preaction={fill, yellow}, pattern=grid] (v1) at (12,6) {};
\node [color=black, preaction={fill, yellow}, pattern=grid] (v2) at (14,6) {};
\node [color=black, preaction={fill, yellow}, pattern=grid] (v3) at (16,6) {};
\node [color=black, preaction={fill, yellow}, pattern=grid] (v4) at (18,6) {};
\node [color=black, preaction={fill, yellow}, pattern=grid] (v5) at (20,6) {};
\node [fill=blue, color=blue] (v0) at (16,8) {};
\node [color=black, preaction={fill, yellow}, pattern=grid] (w1) at (18.5,-5) {};
\node [color=black, preaction={fill, yellow}, pattern=grid] (w2) at (20,-4) {};
\node [color=black, preaction={fill, yellow}, pattern=grid] (w3) at (21.5,-3) {};
\node [color=black, preaction={fill, yellow}, pattern=grid] (w4) at (23,-2) {};
\node [color=black, preaction={fill, yellow}, pattern=grid] (w5) at (24.5,-1) {};
\node [color=black, fill=green] (w0) at (23,-4.5) {};
\draw [very thick] (v1) -- (v2) -- (v3) -- (v4) -- (v5);
\draw [very thick] (u1) -- (u2) -- (u3) -- (u4) -- (u5);
\draw [very thick] (w5) -- (w4) -- (w3) -- (w2) -- (w1);
\draw [very thick]  (v1) edge (v0);
\draw [very thick]  (v2) edge (v0);
\draw [very thick]  (v3) edge (v0);
\draw [very thick]  (v4) edge (v0);
\draw [very thick]  (w4) edge (w0);
\draw [very thick]  (w0) edge (w3);
\draw [very thick]  (w2) edge (w0);
\draw [very thick]  (w1) edge (w0);
\draw [very thick]  (u1) edge (u0);
\draw [very thick]  (u2) edge (u0);
\draw [very thick]  (u0) edge (u3);
\draw [very thick]  (u0) edge (u4);
\draw [very thick]  (u1) edge (v2);
\draw [very thick]  (u2) edge (v1);
\draw [very thick]  (v1) edge (w3);
\draw [very thick]  (w1) edge (v3);
\draw [very thick]  (u1) edge (w3);
\draw [very thick]  (u3) edge (w1);
\draw [very thick]  (u1) edge (w4);
\draw [very thick]  (u4) edge (w1);
\node [color=white, fill=white] (v6) at (2.5,1.5) {};
\node [color=white, fill=white] (v7) at (6.6,1.5) {};
\draw [very thick, -triangle 60] (v6) -- (v7);
\end{tikzpicture}}
	\end{center}
	\caption{
		The reduction described in Lemma~\ref{lem_edge_red}.
		In the example,
		$t = 3$, $t' = \tprim = 4$,
		$p(\mbox{red}) = (1, 2)$,
		$p(\mbox{green}) = (1, 3)$ and
		$p(\mbox{blue}) = (1, 4)$.}
  \label{fig1}
\end{figure}
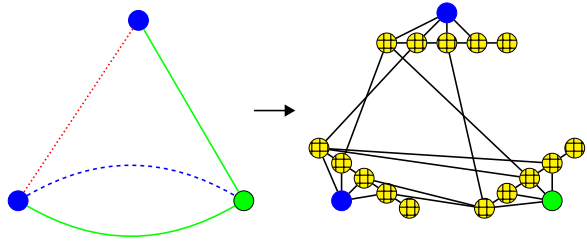
\Proofsketch
    Let $(G, H)$ be an instance of \SIp{$(c, t)$}
    such that $|V(G)| = |V(H)|$.
    Assume that none of the vertices of the instance
    $(G, H)$ was colored yellow.
    Let $t' := 2\left\lceil \sqrt{t} \right\rceil$.
    Note that for $t \geq 1$ we have
    $\tprimhalf \geq 1$ and then
    \[
      {t' \choose 2}
        = \frac{\tprim \cdot (\tprim - 1)}{2}
        \geq \tprimhalf^2
        \geq t.
    \]
    Therefore for each color $x \in [t]$ we can
    pick a different pair $p(x) := (i, j)$ where
    $1 \leq i < j \leq t'$.

    For every vertex $u$ in either the pattern or the host graph, we replace it by a gadget consisting of $t' + 2$ vertices (see Fig.~\ref{fig1}):
    \begin{itemize}
        \item a \emph{center vertex} $u_0'$ of the same color as $u$, and
        \item a path on $t' + 1$ yellow vertices $u'_1, \ldots, u'_{t' + 1}$, the first $t'$ of which are connected to the center vertex.
    \end{itemize}
    For every edge $(u, v)$ of color $x$ in either the pattern or the host graph, we replace it by the edges $(u'_i, v'_j)$ and $(u'_j, v'_i)$ in the modified graph,
    where $(i, j) = p(x)$.
    We denote this new instance of \SIp{$(c + 1, 1)$}
    as $(G', H')$.
    Note that $|V(G')| = (t' + 2) \cdot |V(G)|$
    and $|V(H')| = (t' + 2) \cdot |V(H)|$
    hence $|V(G')| = |V(H')|$ and also
    $|V(G')| = \Oh(\sqrt{t}) \cdot |V(G)|$
    and $|V(H')| = \Oh(\sqrt{t}) \cdot |V(H)|$.
    In Appendix~\ref{app:removing_colors} we
    show that $G$ is a subgraph of $H$ iff $G'$ is
    a subgraph of $H'$.
\QED

Having reduced the number of edge colors down to one,
it remains to reduce the number of vertex colors.
Note that in the following lemma it would be enough to assume $t=1$,
however we prove the lemma in a more general form as it does not 
affect the complexity of the proof.

\begin{lemma}
    \label{lem_vertex_red} $(\spadesuit)$
    An instance $(G, H)$ of \SIp{$(c,t)$}
    such that $|V(G)| = |V(H)|$
    can be reduced to an instance $(G', H')$
    of \SIp{$(1, t)$} with $\Oh(c)$ times
    more vertices such that $|V(G')| = |V(H')|$.
\end{lemma}
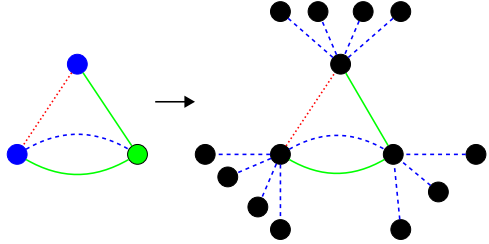
\begin{figure}
\begin{center}
	\scalebox{0.5}{\begin{tikzpicture}[scale=0.4]
\tikzstyle{every node}=[circle, draw, color=black, fill=black, inner sep=0pt, minimum width=15pt]
\node[color=blue, fill=blue] (z1) at (-4.5,4) {};
\node[color=blue, fill=blue] (z2) at (-8.5,-2) {};
\node[color=black, fill=green] (z3) at (-0.5,-2) {};
\draw[color=red, very thick, dotted] (z1) edge (z2);
\draw[color=green, very thick]  (z1) edge (z3);
\draw[color=blue, very thick, bend left, dashed]  (z2) edge (z3);
\draw[color=green, very thick, bend right]  (z2) edge (z3);
\node [color=white, fill=white] (v6) at (0,1.5) {};
\node [color=white, fill=white] (v7) at (4,1.5) {};
\draw [very thick, -triangle 60] (v6) -- (v7);
\node (v1) at (9,-2) {};
\node (v2) at (13,4) {};
\node (v3) at (16.5,-2) {};
\draw [color=red, very thick, dotted]  (v1) edge (v2);
\draw [color=green, very thick]  (v3) edge (v2);
\draw [color=green, very thick, bend right]  (v1) edge (v3);
\draw [color=blue, very thick, bend left, dashed]  (v1) edge (v3);
\node (v4) at (9,7.5) {};
\node (v5) at (11.5,7.5) {};
\node (v8) at (14.5,7.5) {};
\node (v9) at (17,7.5) {};
\node (v10) at (4,-2) {};
\node (v11) at (5.5,-3.5) {};
\node (v12) at (7.5,-5.5) {};
\node (v13) at (9,-7) {};
\node (v15) at (19.5,-4.5) {};
\node (v16) at (22,-2) {};
\node (v14) at (17,-7) {};
\draw [color=blue, very thick, dashed] (v2) edge (v4);
\draw [color=blue, very thick, dashed] (v2) edge (v5);
\draw [color=blue, very thick, dashed] (v8) edge (v2);
\draw [color=blue, very thick, dashed] (v2) edge (v9);
\draw [color=blue, very thick, dashed] (v1) edge (v10);
\draw [color=blue, very thick, dashed] (v1) edge (v11);
\draw [color=blue, very thick, dashed] (v1) edge (v12);
\draw [color=blue, very thick, dashed] (v1) edge (v13);
\draw [color=blue, very thick, dashed] (v14) edge (v3);
\draw [color=blue, very thick, dashed] (v15) edge (v3);
\draw [color=blue, very thick, dashed] (v16) edge (v3);
\end{tikzpicture}}
\end{center}
\caption{
	The reduction described in Lemma~\ref{lem_vertex_red}.
	In the example, the green and blue
	colors of the vertices
	represent the numbers $2$ and $3$
	respectively and the blue color of the edges
	represent the number $1$.}
\end{figure}
\label{fig2}
\Proofsketch
    Let $(G, H)$ be an instance of \SIp{$(c, t)$}
    such that $|V(G)| = |V(H)|$.
    Number the vertex colors arbitrarily from $1$ to $c$ and number the edge colors arbitrarily from
    $1$ to $t$.
    We can assume that for every vertex color
    the number of the vertices in this color in $G$
    and in $H$ is the same because otherwise we can
    produce a trivial NO instance as $(G', H')$.
    In both pattern and host graphs, for each vertex $v$, attach $i + 1$ new leaves
    $v_1, v_2, \ldots, v_{i + 1}$
    to it, where $i$ is the color of $v$,
    using edges of color $1$
    (or any fixed color from $1$ to $t$).
    We also denote $v_0 = v$.
    Consider the \SIp{$(1, t)$} instance $(G', H')$
    on the new graphs.
    For every vertex color the number of the
    vertices in that color in $G$ is the same as in
    $H$ and therefore the number of added leaves is the
    same in $G'$ as in $H'$.
    Hence $|V(G')| = |V(H')|$.
    In Appendix~\ref{app:removing_colors} we show $G$ is a subgraph of $H$ iff $G'$
    is a subgraph of $H'$.
\QED

\Proofof{Lemma \ref{lem_slog}}
    The thesis follows directly from consecutive application of Lemmas \ref{lem_edge_red} and \ref{lem_vertex_red}.
\QED

\section{From \GH to \SI}
\label{sec:homo}

\defproblemu{\GH}{ undirected graphs $G$, $H$.} {Is there a homomorphism from $G$ to $H$, i.e., does there exist a  function $h : V(G) \to V(H)$,
such that for each edge $uv \in E(G)$ we have $h(u)h(v) \in E(H)$.}

In this section we present a reduction which shows that one
can solve the \GH problem by solving $2^{|V(G)| + |V(H)|}$ instances
of the \SI problem, demonstrating
that the lower bound 
of $2^{\Omega(|V(G)| \log V(H) / \log \log V(H))}$ of Fomin et al.~\cite{fomin-homo}
implies an $2^{\Omega(n \log n / \log \log n)}$ lower bound under the Exponential
Time Hypothesis for the \SI problem, where $n = |V(G)| + |V(H)|$.

\Proofof{Theorem~\ref{thm:red-homo}}
Let $(G,H)$ be an instance of \GH and denote $n = V(G) + V(H)$.
Note that any homomorphism $h$ from $G$ to $H$
can be associated with some sequence of non-negative
numbers $(|h^{-1}(v)|)_{v \in V(H)}$, being
the numbers of vertices of $G$ mapped to particular 
vertices of $H$.
The sum of the numbers in such a sequence equals exactly $|V(G)|$.
As the number of such sequences is $\binom{V(G) + V(H)-1}{V(H)-1} \le 2^{n}$,
we can enumerate all such sequences in time $2^n\poly{n}$.
For each such sequence $(a_v)_{v \in V(H)}$
we create a new instance $(G',H')$ of \SI,
where the pattern graph remains the same, i.e., $G' = G$,
and in the host graph $H'$ each vertex of $v \in V(H)$
is replicated exactly $a_v$ times (possibly zero).
Observe that $|V(H')| = |V(G')|$.

We claim that $G$ admits a homomorphism to $H$ iff
for some sequence $(a_v)_{v \in V(H)}$ the
graph $G'$ is a subgraph of $H'$.
First, assume that $G$ admits a homomorphism $h$ to $H$.
Consider the instance $(G',H')$ created
for the sequence $a_v = |h^{-1}(v)|$
and observe that we can create a bijection $h' : V(G') \to V(H')$
by assigning $v \in V(G')$ to its private copy of $h(v)$.
As $h$ is a homomorphism, so is $h'$, and as $h'$ is at the same
time a bijection, we infer that $G'$ is a subgraph of $H'$.

On the other hand if for some sequence $(a_v)_{v\in V(H)}$
the constructed graph $G'$ is a subgraph of $H'$,
then projecting the witnessing injection $g : V(G') \to V(H')$
so that $g'(v)$ is defined as the prototype of the copy $g(v)$
gives a homomorphism from $G$ to $H$,
as copies of each $v \in V(H)$ form independent sets in $H'$.
\QED

\newpage

\bibliographystyle{abbrv}
\bibliography{subiso}

\appendix

\section{Missing proofs from Section~\ref{sec:grouping}}
\label{app:grouping}

Before we prove Lemma~\ref{lem_balanced_coloring}, we show the existence of a potentially unbalanced $9$-coloring.

\begin{lemma}
\label{lem_coloring}
Given a $\TSAT$ formula $\varphi$ with $n$ variables, where each variable occurs in at most $4$ clauses,
we can color the variables of $\varphi$ in polynomial time using at most $9$ colors, 
so that no clause contains two variables of the same color.
\end{lemma}

\Proof
Construct an auxiliary graph $G_\varphi$, the vertex set of which is the set of variables of $S$, 
where two vertices of $G_\varphi$ are adjacent iff they both appear in at least one of the clauses of $\varphi$.
Note that the maximum degree of $G_\varphi$ is bounded by $8$, as each variable appears in at most $4$
clauses and each clause contains at most $3$ literals.
Consequently, we can color $G_\varphi$ with at most $9$ colors in a greedy manner.
\QED

\Proofof{Lemma~\ref{lem_balanced_coloring}}
First, color the variables into $9$ colors using Lemma \ref{lem_coloring}.
Then, while there exists a color with more than $\lceil n / (k - 9) \rceil$ variables assigned to it, 
separate $\lceil n / (k - 9) \rceil$ of them to form a new color.
This can occur at most $k-9$ times, and the lemma follows.
\QED

\section{Missing proofs from Section~\ref{sec:reduction}}
\label{app:reduction}

\Proofof{Claim~\ref{claim:2}}
Let $g$ be a mapping from $P$ to $H_s$ witnessing the fact that
$P$ is a subgraph of $H_s$.
As $g$ respects colors, we infer that the single green
vertex in $P$ is mapped to the single green vertex in $H_s$.
Similarly all the red vertices of $P$ have to be mapped to 
red vertices of $H_s$. 
Additionally the distance between each red vertex $v$
and the green vertex in $P$ cannot be smaller
than the distance between $g(u)$ and the green vertex in $H_s$.
As red vertices induce a path, and the green vertex
is pendant to one if its ends, we infer
that $g$ assigns each vertex of the satisfying-assignment-gadget
in $P$ to its copy in $H_s$ (in short, by construction
there are no non-trivial automorphisms of the gadget).

Construct an assignment $\val : \Var(\varphi) \to \{\true,\false\}$ 
as follows. For a variable $x \in \Var(\varphi)$ find
any clause $C$ that contains $x$
and assign $\val(x)$ to $\true$ iff $g(h(C))_{l(x)} = 1$,
where $h(C)$ is the blue vertex associated with $C$
and $l(x)$ is the color of the variable $x$.
Note that by construction the assignment $\val$ is well-defined,
as edges between blue vertices guarantee consistence.
Consider a clause $C$.
The edges between $h(C)$ and red vertices in the pattern graph $P$
have to be preserved by $g$, and we already observed that
$g$ maps red vertices of $P$ to their corresponding copies in $H_s$.
Hence, we infer that there is an edge between $g(h(C))$ 
and the red vertex $v=(i_1,i_2,i_3,b_1,b_2,b_3)$,
where $b_1,b_2,b_3$ is the only assignment 
to variables of $C$, where $l(C)=\{i_1,i_2,i_3\}$,
which does not satisfy $C$.
This in turn implies that for at least one variable
of $C$ the assignment $\val$ assigns a different
value than the one corresponding to the appropriate bit from $\{b_1,b_2,b_3\}$.
Consequently, $\val$ is a satisfying assignment.
\cqedsymbol

\section{Missing proofs from Section~\ref{sec:slog}}
\label{app:removing_colors}

Here, we present the missing parts of the proof
of Lemmas~\ref{lem_edge_red} and~\ref{lem_vertex_red} from Section~\ref{sec:slog}.

\Proofof{Lemma~\ref{lem_edge_red}}
    It remains to prove that $G$ is a subgraph of $H$ iff $G'$ is
    a subgraph of $H'$.
    If $G$ is a subgraph of $H$ then there exists an
    injective function $f: V(G) \to V(H)$ such that
    edges and colors are preserved.
    Note that every vertex of the instance $(G', H')$
    is of the form $u'_i$ for some vertex $u$
    of the instance $(G, H)$.
    Let $g: V(G') \to V(H')$ be a function such that
    $g(u_i') = f(u)_i'$.
    The function $g$ is an injection because
    the function $f$ is an injection.
    The function $g$ preserves the colors of the vertices
    because
    $col(g(u_0')) = col(f(u)_0') = col(f(u)) = col(u)
      = col(u_0')$
    and for $i > 0$ the color of $u_i'$ is always yellow.
    The function $g$ preserves also the edges.
    Let $u_i' v_j'$ be an edge in $G'$
    such that $i \leq j$ (we can assume this w.l.o.g.).
    If $u = v$ then there exists also an edge
    $f(u)_i' f(u)_j' = g(u_i') g(v_j')$ in $H'$
    because all the gadgets have exactly the same
    structure of the internal edges i.e. if there
    exists an edge $u_i' u_j'$ in a gadget for any
    vertex $u$ then for every vertex $v$
    there exists an edge $v_i' v_j'$
    in a gadget for vertex $v$. 
    If $u \neq v$
    then $1 \leq i < j \leq t'$ and there exists an edge
    $uv$ of the color $x = p^{-1}(i, j)$ in $G$
    and then there exists an edge $f(u)f(v)$ of the
    color $x$ in $H$
    and (because $(i, j) = p(x)$)
    we know that there exists an edge
    $f(u)'_i f(v)_j' = g(u_i') g(v_j')$
    in $H'$.
    The edges of $(G', H')$ have only one color thus
    $g$ preserves the colors of the edges trivially.
    Hence $G'$ is a subgraph of $H'$.
    
    If $G'$ is a subgraph of $H'$ then there exists an
    injective function $g: V(G') \to V(H')$ such that
    edges and colors are preserved.
    The vertices of the form $u_0'$
    are the only vertices of $G'$
    and $H'$ which are not yellow.
    Therefore if $g(u_i') = v_j'$, then $i = 0$ iff $j = 0$.
    If $g(u_0') = v_0'$ then for every $u_i'$
    such that $1 \leq i \leq t'$ we have $g(u_i') = v_j'$
    for some $1 \leq j \leq t'$ because the vertices
    $u_1', u_2', \ldots, u_{t'}'$ are yellow neighbors of
    $u_0'$ and the vertices $v_1', v_2', \ldots, v_{t'}'$
    are the only yellow neighbors of $v_0'$. 
    On the other hand we know that $|V(G)| = |V(H)|$
    and then the number of the vertices of the form
    $u'_i$ for $1 \leq i \leq t'$ is the same in
    $G'$ and in $H'$.
    Therefore
    for every vertex $v_i'$ in $H'$ such that
    $1 \leq i \leq t'$ there exists a vertex $u_j'$
    in $G'$ such that $1 \leq j \leq t'$ and
    $g(u_j') = v_i'$. 
    Therefore if $g(u_i') = v_j'$ then $i = t' + 1$ iff
    $j = t' + 1$.
    Moreover the vertices $u_1', u_2', \ldots, u_{t'}'$
    create a path (in this order)
    and the only directed paths
    containing exactly the vertices
    $\{v_1', v_2', \ldots, v_{t'}'\} =
      g(\{u_1', u_2', \ldots, u_{t'}'\})$ are
    $v_1', v_2', \ldots, v_{t'}'$ and
    $v_{t'}', v_{t' - 1}', \ldots, v_1'$.
    But the vertex $u_{t'}'$ is a neighbor of
    the vertex $u_{t' + 1}'$
    and the vertex $v_1'$ has no neighbor of the form
    $w_{t' + 1}'$ for any vertex $w$ in $H$.
    On the other hand the vertex
    $u_{t' + 1}'$ has to be mapped to
    a vertex of the form $w_{t' + 1}'$ for some
    vertex $w$ in $H$.
    Therefore the path $u_1', u_2', \ldots, u_{t'}'$
    is mapped to the path $v_1', v_2', \ldots, v_{t'}'$
    i.e. for every vertex $u_i'$
    such that $1 \leq i \leq t'$
    we have $g(u_i') = v_i'$.
    Let $f: V(G) \to V(H)$ be a function such that
    $f(u) = v$ iff $g(u_0') = v_0'$.
    (note that then $f(u)_0' = v_0' = g(u_0')$).
    The function $f$ is an injection because the function
    $g$ is an injection.
    The function $f$ preserves the colors of the vertices
    because
    $col(f(u)) = col(f(u)_0')
      = col(g(u_0')) = col(u_0') = col(u)$.
    The function $f$ preserves also the edges with their
    colors because if there is an edge $uv$ of the color
    $x$ in the graph $G$ then for $(i, j) = p(x)$
    there is an edge $u_i' v_j'$ in the graph $G'$
    and therefore there is an edge
    $g(u_i') g(v_j') = f(u)_i' f(v)_j'$
    in the graph $H'$
    and then
    there is an edge $f(u) f(v)$ of the color $x$
    in the graph $H$.
    Hence $G$ is a subgraph of $H$.
\QED
    
\Proofof{Lemma~\ref{lem_vertex_red}}
    If $G$ is a subgraph of $H$ then there exists an
    injective function $f: V(G) \to V(H)$ such that
    edges and colors are preserved.
    Note that every vertex of the instance $(G', H')$
    is of the form $v_i$ for some vertex $v$
    of the instance $(G, H)$.
    Let $g: V(G') \to V(H')$ be a function
    such that $g(v_i) = f(v)_i$ which is a correctly
    defined function
    because $col(v) = col(f(v))$ and therefore
    $v_0$ has the same number of leaves
    in the graph $G'$
    as $f(v)_0$ in the graph $H'$.
    The function $g$ is an injection because the
    function $f$ is an injection.
    The function $g$ preserves the colors of the vertices
    trivially.
    We show that the function $g$
    preserves also edges and their colors.
    Let assume that there is an edge $u_i v_j$ for
    $i \leq j$ (we can assume that w.l.o.g)
    of the
    color $x$ in the graph $G'$.
    If $j > 0$
    then $u = v$, $i = 0$ and $x = 1$
    and there exists also an edge
    $f(u)_0 f(u)_j = g(u_i) g(v_j)$ of the color
    $1 = x$ in the graph $H'$.
    Otherwise we have $i = j = 0$ and then there exists
    an edge $uv$ of the color $x$ in the graph $G$
    thus there exists an edge $f(u)f(v)$ of the color $x$
    in the graph $H$ hence there exists an edge
    $f(u)_0 f(v)_0 = g(u_i) g(v_j)$ of the color $x$
    in the graph $H'$.
    Therefore $G'$ is a subgraph of $G$.
    
    If $G'$ is a subgraph of $H'$ then there exists an
    injective function $g: V(G') \to V(H')$ such that
    edges and colors are preserved.
    All vertices from the original pattern graph have to be matched to vertices from the original host graph, as they are the only ones of degree greater than $1$ in the new graphs.
    But the number of the vertices of the form $u_0$
    is the same in $G'$ as in $H'$ because
    $|V(G)| = |V(H)|$.
    Therefore for every
    vertex of the form $v_0$ in $H'$ there exists
    a vertex of the form $u_0$ in $G'$ such that
    $g(u_0) = v_0$.
    Hence, the leaves have to map to leaves.
    But the number of leaves is the
    same in $G'$ as in $H'$.
    Thus for every leaf $v_i$ in $H'$ there exists
    a leaf $u_j$ in $G'$ such that $g(u_j) = v_i$.
    Hence all the leaves are used and then for every
    vertex $u_0$ in $G'$ the number of leaves of
    $u_0$ in $G'$ is the same as the number of leaves
    of $g(u_0)$ in $H'$.
    Let us consider a function $f: V(G) \to V(H)$
    such that $f(u) = v$ iff $g(u_0) = v_0$
    (then $f(u)_0 = v_0 = g(u_0)$).
    Note that $f = g|_{V(G)}$.
    The function $f$ is an injection because the function
    $g$ is an injection.
    The function $f$ preserves the colors of the vertices
    because for every $v$ in the graph $G$ we have that
    $v_0$ has exactly $col(v) + 1$ leaves as neighbors
    in the graph $G'$
    and then $g(v_0) = f(v)_0$ has also exactly
    $col(v) + 1$ leafs as neighbors.
    But on the other hand
    the vertex $f(v)_0$ has exactly
    $col(f(v)) + 1$ leafs as neighbors
    in the graph $H'$
    and therefore $col(f(v)) = col(v)$.
    The function $f$ preserves also the edges with
    their colors because for every edge $uv$ of a color
    $x$ in the graph $G$ there exists an edge $u_0 v_0$
    of the color $x$ in the graph $G'$
    and therefore there exists an edge
    $g(u_0) g(v_0) = f(u)_0 f(v)_0$ of the color $x$
    in the graph $H'$ and
    hence there exists an edge $f(u) f(v)$ of the color
    $x$ in the graph $H$.
    Therefore $G$ is a subgraph of $H$.
\QED

\end{document}